\documentclass[
 aip,showpacs,superscriptaddress,numerical,amsmath,amssymb,floatfix,floatfix,
reprint,
]{revtex4-1}
\usepackage{xcolor}
\usepackage[
	pdffitwindow=true,
	colorlinks=true,
	frenchlinks=false,
        linkcolor=blue,
	anchorcolor=blue,
        citecolor=blue,
        filecolor=blue,
        urlcolor=blue,
        bookmarks=true,
        bookmarksopen=true,
	bookmarksnumbered=true,
        bookmarksopenlevel=1,
        plainpages=false,
	pdfpagelayout=TwoPageLeft,
        pdfpagelabels=true,
	breaklinks
]{hyperref}
\usepackage[per-mode=symbol,separate-uncertainty]{siunitx}
\usepackage{graphicx}
\usepackage{dcolumn}
\usepackage{bm}
\usepackage[english]{babel}
\usepackage{color}
\usepackage{ulem}
\usepackage{todonotes}

\begin{document}
\title[]{Quantitative comparison of magnon transport experiments in three-terminal YIG/Pt nanostructures acquired via dc and ac detection techniques}

\author{J.~G{\"u}ckelhorn}
\email[]{janine.gueckelhorn@wmi.badw.de}
\affiliation{Walther-Mei{\ss}ner-Institut, Bayerische Akademie der Wissenschaften, 85748 Garching, Germany}
\affiliation{Physik-Department, Technische Universit\"{a}t M\"{u}nchen, 85748 Garching, Germany}
\author{T.~Wimmer}
\affiliation{Walther-Mei{\ss}ner-Institut, Bayerische Akademie der Wissenschaften, 85748 Garching, Germany}
\affiliation{Physik-Department, Technische Universit\"{a}t M\"{u}nchen, 85748 Garching, Germany}
\author{S.~Gepr{\"a}gs}
\affiliation{Walther-Mei{\ss}ner-Institut, Bayerische Akademie der Wissenschaften, 85748 Garching, Germany}
\author{H.~Huebl}
\affiliation{Walther-Mei{\ss}ner-Institut, Bayerische Akademie der Wissenschaften, 85748 Garching, Germany}
\affiliation{Physik-Department, Technische Universit\"{a}t M\"{u}nchen, 85748 Garching, Germany}
\affiliation{Munich Center for Quantum Science and Technology (MCQST), Schellingstr. 4, D-80799 M\"{u}nchen, Germany}
\author{R.~Gross}
\affiliation{Walther-Mei{\ss}ner-Institut, Bayerische Akademie der Wissenschaften, 85748 Garching, Germany}
\affiliation{Physik-Department, Technische Universit\"{a}t M\"{u}nchen, 85748 Garching, Germany}
\affiliation{Munich Center for Quantum Science and Technology (MCQST), Schellingstr. 4, D-80799 M\"{u}nchen, Germany}
\author{M.~Althammer}
\email[]{matthias.althammer@wmi.badw.de}
\affiliation{Walther-Mei{\ss}ner-Institut, Bayerische Akademie der Wissenschaften, 85748 Garching, Germany}
\affiliation{Physik-Department, Technische Universit\"{a}t M\"{u}nchen, 85748 Garching, Germany}

\date{\today}


\begin{abstract}

All-electrical generation and detection of pure spin currents is a promising way towards controlling the diffusive magnon transport in magnetically ordered insulators. We quantitatively compare two measurement schemes, which allow to measure the magnon spin transport in a three-terminal device based on a yttrium iron garnet thin film.
We demonstrate that the dc charge current method based on the current reversal technique and the ac charge current method utilizing first and second harmonic lock-in detection can both efficiently distinguish between electrically and thermally injected magnons. In addition, both measurement schemes allow to investigate the modulation of magnon transport induced by an additional dc charge current applied to the center modulator strip. However, while at low modulator charge current both schemes yield identical results, we find clear differences above a certain threshold current. This difference originates from nonlinear effects of the modulator current on the magnon conductance.  

\end{abstract}%
\maketitle
In the field of spintronics, pure spin currents are promising for spin and information transport at low dissipation level. To this end, the efficient control of pure spin currents is an essential, but challenging task.~\cite{Ahopelto2019, Jansen2012, Hoffmann2007, Coll2019} In magnetically ordered insulators, spin currents are carried by magnons, the elementary excitations of the spin system. These magnonic spin currents lead to interesting new device concepts for magnon-based information processing.~\cite{Chumak2015magnon,Chumak_2017,Nakata_2017,LudoTransistor} In this context, devices for magnon logic operations mainly focus on coherent magnon transport. For instance, it has been shown that damping compensation via spin transfer torque is an efficient method to optimize coherent magnon propagation.~\cite{An2014,Gladii2016,Demidov2011,Demidov2014} Furthermore, a logic majority gate~\cite{Chumak2017majoritygate} and the first magnon transistor~\cite{Chumak2014magnontransistor} using magnonic crystals~\cite{Krawczyk_2014} have been implemented. 

\begin{figure}
	\includegraphics[]{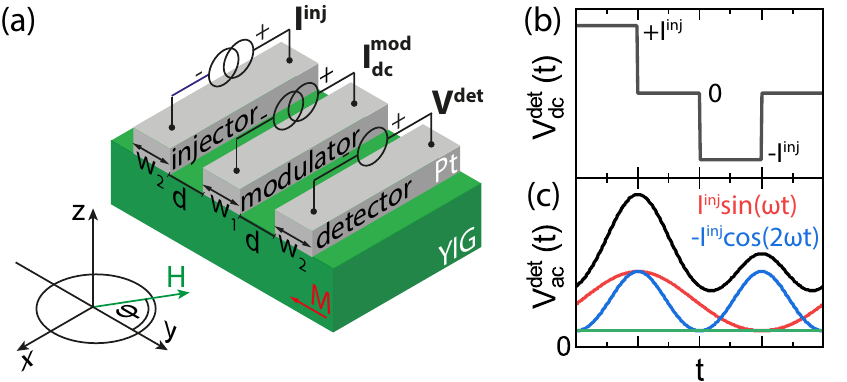}
	\caption[]{(a) Sketch of the sample configuration with the electrical wiring scheme and the electrical connection scheme, and the coordinate system with the in-plane rotation angle $\varphi$ of the applied magnetic field $\mu_0\mathbf{H}$. (b), (c) Schematic dependence of the detector voltage $V^\mathrm{det}$ as a function of time according to Eq.~(\ref{eq:Vdet}) for (b) the dc and (c) the ac technique. (b) For the dc technique, the current to the injector $I^\mathrm{inj}$ is stepwise varied from $+I^\mathrm{inj}$ to $-I^\mathrm{inj}$ and vice versa. (c) For the ac technique, $V^\mathrm{det}_\mathrm{ac}$ is shown for the first (red) and second (blue) harmonic signal as well as for a constant offset detector voltage (green). The black line corresponds to their superposition.}
	\label{fig:fig1device}
\end{figure}

Recently, incoherent, thermally excited magnons have gained increasing interest as information carriers for logic operations. In bilayer systems consisting of magnetically ordered insulators (MOI) and heavy metals (HM) with strong spin orbit coupling, it has been shown that incoherent magnons in the MOI can be excited electrically~\cite{CornelissenMMR,SchlitzMMR,Casanova2016} as well as thermally~\cite{CornelissenMMR,GilesSSE, VanWeesThickness}, which then can be detected electrically in the HM utilizing the inverse spin Hall effect (SHE).~\cite{HirschSHE,Dyakonov,Althammer2018} Moreover, devices based on non-continuous HM electrodes have been used to show that a superposition of diffusive magnon currents allow for the realization of a majority gate.~\cite{KathrinLogik} Later on, similar device concepts were used to demonstrate the manipulation of magnon currents using a three-electrode arrangement in yttrium iron garnet ($\mathrm{Y_3Fe_5O_{12}}$, YIG)/Pt bilayers.~\cite{LudoTransistor,Wimmer2019spin,Klaui2019,Das2020} In these experiments, a charge current is applied to a Pt strip (injector) injecting magnons into the YIG via the SHE and Joule heating (see Fig.~\ref{fig:fig1device}(a)). These magnons are then electrically detected via the inverse SHE as a voltage signal at a second Pt strip (detector). A charge current applied to a third Pt strip (modulator) placed between these two Pt strips allows to manipulate the magnon transport from injector to detector.~\cite{LudoTransistor}
The effect of the modulator in these experiments can be modeled as a change in the effective magnon conductivity, which has to be distinguished from the expected change in the magnon transport signature due to spin Hall and spin Seebeck physics. In particular, the effective magnon resistance changes in a nonlinear fashion with the modulator current and shows a threshold behavior.
Two main measurement schemes have been used to access the magnon transport properties, which are based on an ac~\cite{CornelissenMMR, LudoTransistor, Wimmer2019spin} and a dc~\cite{SchlitzMMR, KleinMMR, Klaui2019} stimulus applied to the injector. Although it is not obvious whether or not these techniques yield exactly the same result, a quantitative comparison is still missing.

In this paper, we perform a quantitative comparison of the following measurement schemes: (i) a \textit{dc-detection technique} utilizing the current reversal method~\cite{SchreierCSSE} and (ii) an \textit{ac-readout technique} based on lock-in detection.
We corroborate that both techniques are quantitatively equivalent in the regime where the magnon resistance is weakly affected by the modulator current. In the nonlinear regime we find that the two techniques are qualitatively different which gives access to higher order terms originating from the injector current.

As shown in Fig.~\ref{fig:fig1device}(a), we investigate the magnon transport using a three-terminal YIG/Pt nanostructure.~\cite{LudoTransistor}
A charge current $I^\mathrm{inj}$ is applied to the Pt-injector, inducing a magnon accumulation in the YIG film, both via the SHE generated spin accumulation and via Joule heating. The magnons diffuse to the Pt-detector strip, where they induce a voltage $V^\mathrm{det}$ via the inverse SHE. A dc charge current $I^\mathrm{mod}_\mathrm{dc}$ applied to along the Pt-modulator allows to manipulate the magnon transport between injector and detector via a SHE induced spin accumulation and Joule heating effects. 

Following previous works~\cite{LudoTransistor,Wimmer2019spin,CornelissenTheory,KleinMMR}, we express the detector voltage as
\begin{equation}
V^\mathrm{det}\left(I^\mathrm{inj},I^\mathrm{mod}\right) =  \sum_{i\in\left\lbrace \mathrm{inj,mod}\right\rbrace }\sum_{j=1}^\infty R_j^{i\mathrm{\text{-}det}}\left(I^\mathrm{mod}\right)\cdot\left[{I^{i}}\right]^j.
\label{eq:Vdet}
\end{equation}
Here, $R_{j}^{i\mathrm{\text{-}det}}\left( I^\mathrm{mod}\right)$ are the transport coefficients describing the conversion process at the YIG/Pt interface and the transport in the YIG layer. Note that we only account for changes in $R_{j}^{i}$ via $I^\mathrm{mod}$. This assumption is only valid for small injector currents.~\cite{LudoTransistor,Wimmer2019spin}

For the dc-detection technique, we utilize an advanced current reversal scheme. We apply a dc charge current sequence $+I^\mathrm{inj},0, -I^\mathrm{inj}$ to the injector, while a constant charge current $I^\mathrm{mod}_\mathrm{dc}$ is applied to the modulator, and measure for each configuration the voltage $V^\mathrm{det}_\mathrm{dc}$ at the detector, as sketched in Fig.~\ref{fig:fig1device}(b). From these measurements, we can then define
\begin{equation}
\begin{split}
V&^\mathrm{SHE}_\mathrm{dc} = \frac{1}{2}\left[  V^\mathrm{det}_\mathrm{dc}\left(I^\mathrm{inj},I^\mathrm{mod}_\mathrm{dc}\right)-V^\mathrm{det}_\mathrm{dc}\left(-I^\mathrm{inj},I^\mathrm{mod}_\mathrm{dc}\right)\right]  \\
&=  R_1^{\mathrm{inj\text{-}det}}\left( I^\mathrm{mod}\right)I^\mathrm{inj} + 
R_3^{\mathrm{inj\text{-}det}}\left( I^\mathrm{mod}\right){I^\mathrm{inj}}^3+...
\end{split}
\label{eq:VSHE}
\end{equation}
as the voltage due to the SHE induced magnons 
transported from the injector to the detector 
assuming an odd symmetry with respect to $I^\mathrm{inj}$. In similar fashion, we define
\begin{equation}
\begin{split}
V&^\mathrm{therm}_\mathrm{dc}= \frac{1}{2}\left[ V^\mathrm{det}_\mathrm{dc}\left(I^\mathrm{inj},I^\mathrm{mod}_\mathrm{dc}\right)+V^\mathrm{det}_\mathrm{dc}\left(-I^\mathrm{inj},I^\mathrm{mod}_\mathrm{dc}\right)\right.\\
&\left.-2V^\mathrm{det}_\mathrm{dc}\left(0,I^\mathrm{mod}_\mathrm{dc}\right)\right] = R_2^{\mathrm{inj\text{-}det}}\left( I^\mathrm{mod}\right){I^\mathrm{inj}}^2+...
\end{split}
\label{eq:Vtherm}
\end{equation}
as the voltage due to the thermally injected magnons assuming an even symmetry with respect to $I^\mathrm{inj}$. 
This elaborate scheme allows us to disentangle the dc detector voltages generated by $I^\mathrm{mod}$ and $I^\mathrm{inj}$. Thus, $V^\mathrm{SHE}_\mathrm{dc}$ and $V^\mathrm{therm}_\mathrm{dc}$ only contain contributions from $I^\mathrm{mod}$ via the transport coefficients $R_{j}^\mathrm{inj\text{-}det}\left( I^\mathrm{mod}\right)$.

In case of the ac-readout technique, we simultaneously apply an ac charge current $I^\mathrm{inj}_\mathrm{ac}(t)=I^\mathrm{inj}\sin(\omega t)$ to the injector and a constant dc charge current $I^\mathrm{mod}_\mathrm{dc}$ to the modulator and record the first and second harmonic signal of $V^\mathrm{det}_\mathrm{ac}$ via lock-in detection (compare Fig.~\ref{fig:fig1device}(c)). For the first harmonic signal $V^\mathrm{1\omega}_\mathrm{ac}$ and a time interval $T\gg 1/\omega$ we obtain:
\begin{equation}
\begin{split}
&V^\mathrm{1\omega}= \frac{2}{T}\int_{0}^{T}\sin(\omega t)V^\mathrm{det}_\mathrm{ac}\left(I^\mathrm{inj}_\mathrm{ac}(t), I^\mathrm{mod}_\mathrm{dc}\right)dt \\
&=  R_1^{\mathrm{inj\text{-}det}}\left( I^\mathrm{mod}\right)I^\mathrm{inj} 
+  \frac{3}{4}R_3^{\mathrm{inj\text{-}det}}\left( I^\mathrm{mod}\right){I^\mathrm{inj}}^3+...
\end{split}
\label{eq:AC1}
\end{equation} 
which corresponds to the SHE induced magnon transport signal. For the second harmonic signal $V^\mathrm{2\omega}_\mathrm{ac}$ we obtain:
\begin{equation}
\begin{split}
V&^\mathrm{2\omega}_\mathrm{ac} = -\frac{2}{T}\int_{0}^{T}\cos(2\omega t)V^\mathrm{det}_\mathrm{ac}\left(I^\mathrm{inj}_\mathrm{ac}(t), I^\mathrm{mod}_\mathrm{dc}\right)dt\\
&= \frac{1}{2}R_2^{\mathrm{inj\text{-}det}}\left( I^\mathrm{mod}\right){I^\mathrm{inj}}^2+...
\end{split}
\label{eq:AC2}
\end{equation}
which corresponds to the thermally generated magnons via Joule heating in the injector. When measuring $V^\mathrm{2\omega}_\mathrm{ac}$ one has to account for the $-\SI{90}{\degree}$ phase shift of the signal with respect to the reference signal.
Due to the lock-in technique, the first and second harmonic signal only contain contributions from the magnon transport between injector and detector. 

If we now compare $V^\mathrm{SHE}_\mathrm{dc}$ with $V^\mathrm{1\omega}_\mathrm{ac}$, we see that these two quantities should be identical if $R_j^{\mathrm{inj\text{-}det}} = 0$ for $j\ge2$. Thus, a quantitative comparison of $V^\mathrm{SHE}_\mathrm{dc}$ and $V^\mathrm{1\omega}_\mathrm{ac}$ enables us to obtain information on higher order SHE contributions. In contrast, the ratio $V^\mathrm{2\omega}_\mathrm{ac}/V^\mathrm{therm}_\mathrm{dc}$ is constant and yields $1/2$ if only transport coefficients up to the fifth order ($j\le5$) contribute. To confirm this model conjecture, we conducted magnon transport experiments in YIG/Pt heterostructures. 

For the experiment comparing the dc- and ac-detection techniques, we use a peak value of $I^\mathrm{inj} = \SI{100}{\micro\ampere}$ and in the case of lock-in detection a low frequency ($\SI{7.737}{\hertz}$) modulation of the ac charge current.
The device consists of  $\SI{5}{\nano\metre}$ thick Pt strips with an edge-to-edge distance of $d=\SI{200}{\nano\metre}$ and a modulator width of $w_1=\SI{500}{\nano\metre}$ on a $\SI{11.4}{\nano\metre}$ thick YIG film (see supplemental information for growth details). The injector and the detector have a width of $w_2=\SI{500}{\nano\metre}$ and a length of $l_2 = \SI{50}{\micro\metre}$, while the length of the modulator is $l_1 = \SI{64}{\micro\metre}$. 

\begin{figure}
	\includegraphics[]{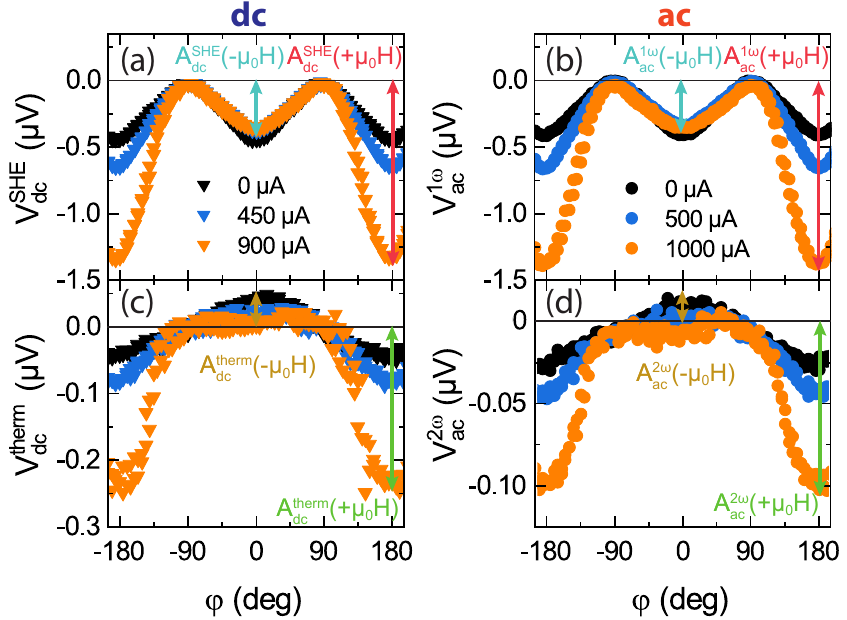}
	\caption[]{Detector signals (a) $V^\mathrm{SHE}_\mathrm{dc}$, (b) $V^\mathrm{1\omega}_\mathrm{ac}$, (c) $V^\mathrm{therm}_\mathrm{dc}$, (d) $V^\mathrm{2\omega}_\mathrm{ac}$ plotted versus the magnetic field orientation with constant magnitude $\mu_0H = \SI{50}{\milli\tesla}$ for various positive modulator currents $I^\mathrm{mod}_\mathrm{dc}$. For $I_\mathrm{dc}^\mathrm{mod} > 0$, the magnon transport signal is significantly increased at $\varphi = \pm \SI{180}{\degree}$ and reduced at $\varphi =  \SI{0}{\degree}$. For the SHE induced magnon transport signals the (a) dc detector signal $V^\mathrm{SHE}_\mathrm{dc}$ and (b) the first harmonic signal of the ac measurement technique $V^\mathrm{1\omega}_\mathrm{ac}$ are in perfect agreement. While the angle dependence of the thermal signals (c) $V^\mathrm{therm}_\mathrm{dc}$ and (d) $V^\mathrm{2\omega}_\mathrm{ac}$ is in good agreement, their absolute amplitude values strongly differ. The voltage amplitudes $A^\mathrm{SHE}_\mathrm{dc}$, $A^\mathrm{1\omega}_\mathrm{ac}$, $A^\mathrm{therm}_\mathrm{dc}$, $A^\mathrm{2\omega}_\mathrm{ac}$ are extracted from the angle dependence of the detector signals as shown by the vertical arrows.}
	\label{fig:fig2admr}
\end{figure}

To characterize the magnon transport in our device, we plot the voltage signals $V^\mathrm{SHE}_\mathrm{dc}$, $V^\mathrm{therm}_\mathrm{dc}$, $V^\mathrm{1\omega}_\mathrm{ac}$, $V^\mathrm{2\omega}_\mathrm{ac}$ as a function of the magnetic field orientation $\varphi$ (cf. Fig.~\ref{fig:fig1device}(a)) measured with a fixed magnetic field strength of $\mu_0H = \SI{50}{\milli\tesla}$ at $T = \SI{280}{\kelvin}$ for various positive modulator currents $I^\mathrm{mod}_\mathrm{dc}$. We first focus on $V^\mathrm{SHE}_\mathrm{dc}$ and $V^\mathrm{1\omega}_\mathrm{ac}$ in Fig.~\ref{fig:fig2admr}(a) and (b). For  $I^\mathrm{mod}_\mathrm{dc} = 0$ (black data points), we observe the distinctive $\cos^2\varphi$ modulation for magnon transport between the injector and detector for both measurement techniques with minima in $V^\mathrm{SHE}_\mathrm{dc}$ and $V^\mathrm{1\omega}_\mathrm{ac}$ for $\mathbf{H}\parallel\pm\hat{\mathbf{y}}$  ($\varphi=\SI{-180}{\degree},\SI{0}{\degree},\SI{180}{\degree}$), corresponding to maxima in magnon transport between injector and detector.~\cite{CornelissenMMR,SchlitzMMR} For $I_\mathrm{dc}^\mathrm{mod} > 0$, the magnon transport signal is significantly increased at $\varphi = \pm \SI{180}{\degree}$ for $V^\mathrm{SHE}_\mathrm{dc}$ as well as $V^\mathrm{1\omega}_\mathrm{ac}$. This enhancement can be explained as an increase in magnon conductivity due to a magnon accumulation underneath the modulator caused by the SHE induced magnon chemical potential and thermally generated magnons due to Joule heating. This increase in magnon conductivity leads to a larger magnon transport signal at the detector and thus larger negative voltage in both measurement schemes. At $\varphi = \SI{0}{\degree}$, we obtain a decreased magnon transport signal for $V^\mathrm{SHE}_\mathrm{dc}$ as well as $V^\mathrm{1\omega}_\mathrm{ac}$. This originates from the magnon depletion caused by the annihilation of magnons via the SHE. However, this depletion is counterbalanced by the thermally injected magnons arising due to Joule heating of the modulator strip.
Comparing the dc and ac case, not just the angle dependence is equivalent, but also the voltage amplitudes $V_\mathrm{dc}^\mathrm{SHE}$ and $V_\mathrm{ac}^\mathrm{1\omega}$ are in agreement with the predictions from our detector voltage model. Separate measurements on an additional sample yield identical results (see supplementary material). 

We now discuss the angle-dependent data obtained from the thermal signals $V^\mathrm{therm}_\mathrm{dc}$ and $V^\mathrm{2\omega}_\mathrm{ac}$. In Fig.~\ref{fig:fig2admr}(c) and (d) we plot the angle-dependent thermal voltage signals for the dc- and ac-detection technique for positive $I^\mathrm{mod}_\mathrm{dc}$, respectively. For $I_\mathrm{dc}^\mathrm{mod}=0$ the measurements of the thermally induced magnons show the characteristic $\cos\varphi$ modulation in agreement with previous work.~\cite{CornelissenMMR}
For $I_\mathrm{dc}^\mathrm{mod}>0$, we observe a significant increase of the detector signals $V_\mathrm{dc}^\mathrm{therm}$ and $V_\mathrm{ac}^\mathrm{2\omega}$ at  $\varphi = \pm \SI{180}{\degree}$ and a decrease at $\varphi = \SI{0}{\degree}$ as already reported in Ref.~\onlinecite{Wimmer2019spin}. For $I_\mathrm{dc}^\mathrm{mod}=\SI{900}{\micro\ampere}$ and $\SI{1000}{\micro\ampere}$, this difference is significantly increased. We attribute this enhancement and decrease of the signal to the same mechanisms as in the case for the SHE driven magnon transport ($V^\mathrm{SHE}_\mathrm{dc}$ and $V^\mathrm{1\omega}_\mathrm{ac}$). At $\varphi =\pm \SI{180}{\degree}$, the magnon conductance underneath the modulator is increased by the SHE and thermally injected magnons via $I^\mathrm{mod}_\mathrm{dc}$. At  $\varphi = \SI{0}{\degree}$, the magnon depletion underneath the modulator is counterbalanced by the thermally injected magnons and only a small reduction in the signal amplitude is observed. 
Comparing dc and ac configuration, we observe that the thermally induced signals $V_\mathrm{dc}^\mathrm{therm}$ and $V_\mathrm{ac}^\mathrm{2\omega}$ strongly differ in their absolute amplitude values, as expected from our model.

\begin{figure}
	\includegraphics[]{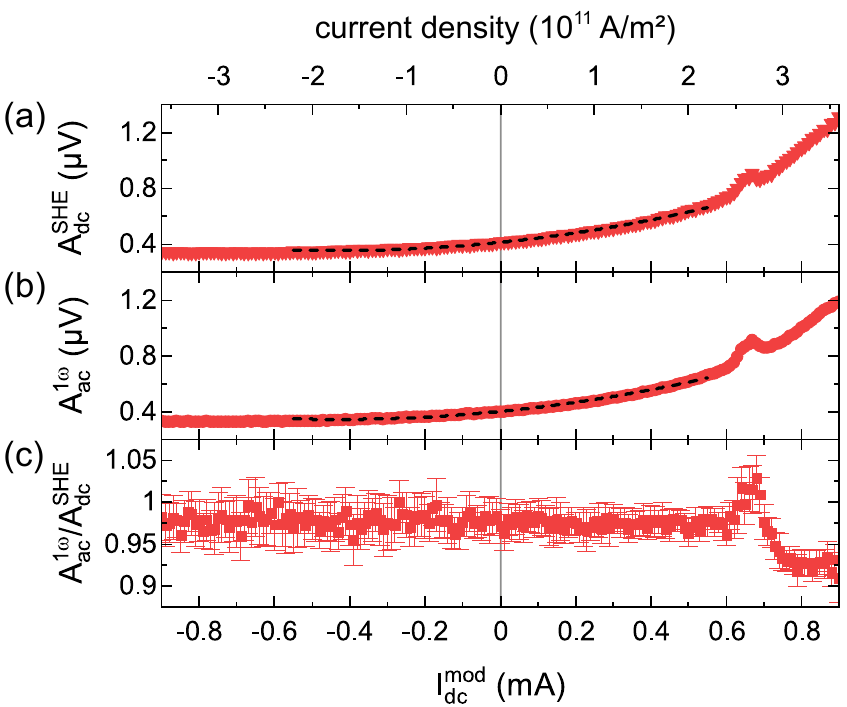}
	\caption[]{Extracted amplitudes (a) $A^\mathrm{SHE}_\mathrm{dc}$ and (b) $A^\mathrm{1\omega}_\mathrm{ac}$ for $\mu_0H=\SI{60}{\milli\tesla}$ (as indicated in Fig.~\ref{fig:fig2admr}) of the SHE injected magnon transport signal versus the dc charge current $I^\mathrm{mod}_\mathrm{dc}$. The curves and signal amplitudes show similar behavior for (a) the dc and (b) the ac scheme. The black dashed line is a fit indicating the $I^\mathrm{mod}_\mathrm{dc}+{I^\mathrm{mod}_\mathrm{dc}}^2$ dependence in the low bias regime ($|I^\mathrm{mod}_\mathrm{dc}|\leq\SI{0.55}{\milli\ampere}$). (c) Ratio of the extracted amplitudes $A^\mathrm{SHE}_\mathrm{dc}$ and $A^\mathrm{1\omega}_\mathrm{ac}$. For $I^\mathrm{mod}_\mathrm{dc} \leq \SI{0.55}{\milli\ampere}$, the ratio shows a nearly constant behavior ($A_\mathrm{ac}^\mathrm{1\omega}/A_\mathrm{dc}^\mathrm{SHE}\simeq\mathrm{0.98}$). For higher modulator current values, the ratio clearly deviates from 0.98.}
	\label{fig:fig4currsweep200nm500nm}
\end{figure}

For a more elaborate quantitative  comparison of the detected voltages in dc and ac measurements, we extract the signal amplitudes $A^\mathrm{SHE}_\mathrm{dc}(\pm\mu_0H)$ and $A^\mathrm{1\omega}_\mathrm{ac}(\pm\mu_0H)$ of the angle-dependent measurements, as indicated in Fig.~\ref{fig:fig2admr}, and plot them as a function of $I^\mathrm{mod}_\mathrm{dc}$ for a magnetic field magnitude of $\mu_0H = \SI{60}{\milli\tesla}$ in Fig.~\ref{fig:fig4currsweep200nm500nm}. 
We note that we use $A^\mathrm{SHE}_\mathrm{dc}$ and $A^\mathrm{1\omega}_\mathrm{ac}$ in our analysis instead of $V^\mathrm{SHE}_\mathrm{dc}$ and $V^\mathrm{1\omega}_\mathrm{ac}$, since at $\varphi=\SI{90}{\degree}$ the voltage measured is close to 0 leading to significant contributions of noise.
At first glance, the curves and the signal amplitudes show similar behavior for the dc (Fig.~\ref{fig:fig4currsweep200nm500nm}(a)) and ac (Fig.~\ref{fig:fig4currsweep200nm500nm}(b)) configuration. As reported in Refs.~\onlinecite{LudoTransistor,Wimmer2019spin}, the signal amplitudes can be modeled by a superposition of a linear (SHE) and quadratic (Joule heating) dependence in the low bias regime ($|I^\mathrm{mod}_\mathrm{dc}|\leq\SI{0.55}{\milli\ampere}$). 
To illustrate this, we plot this linear and quadratic dependence as a black dashed line in Fig.~\ref{fig:fig4currsweep200nm500nm}(a) and (b). The fit well reproduces the measured data points in the low bias regime. 
For larger currents ($I^\mathrm{mod}_\mathrm{dc}>\SI{0.55}{\milli\ampere}$) we observe a pronounced deviation from this behavior. This enhancement in magnon conductance is in agreement with our previous work, which we attribute to a zero effective damping state via SHE induced damping-like spin-orbit torque underneath the modulator.~\cite{Wimmer2019spin} 
To show that the extracted amplitudes as a function of the modulator current $I^\mathrm{mod}_\mathrm{dc}$ for the dc configuration is in accordance with the ac measurement technique, the ratio $A_\mathrm{ac}^\mathrm{1\omega}/A_\mathrm{dc}^\mathrm{SHE}$ is plotted in Fig.~\ref{fig:fig4currsweep200nm500nm}(c).
In the low and negative bias regime ($I^\mathrm{mod}_\mathrm{dc}\leq\SI{0.55}{\milli\ampere}$) the ratio is nearly constant with $A_\mathrm{ac}^\mathrm{1\omega}/A_\mathrm{dc}^\mathrm{SHE}\simeq\mathrm{0.98}$. This value is close to 1, which our model predicts for only linear effects with  $R_j^{\mathrm{inj\text{-}det}}\left(I^\mathrm{mod}_\mathrm{dc}\right)  = 0$ for $j\ge2$. However, for $I^\mathrm{mod}_\mathrm{dc}>\SI{0.55}{\milli\ampere}$ the ratio exhibits a clear deviation from 1. 
Following the arguments of our theoretical model, this deviation indicates that for $I^\mathrm{mod}_\mathrm{dc}>\SI{0.55}{\milli\ampere}$ $R_j^{\mathrm{inj\text{-}det}}\left(I^\mathrm{mod}_\mathrm{dc}\right)\neq0$ (for $j\ge2$), i.e. a deviation from the linear $I^\mathrm{inj}$ dependence. We attribute this to a new regime established via the damping compensation underneath the modulator, reflecting a typical threshold behavior of nonlinear effects.~\cite{KleinMMR}
For negative field polarity we extract a similar dependence of the ratio $A^\mathrm{1\omega}_\mathrm{ac}/A^\mathrm{SHE}_\mathrm{dc}$ just with a threshold for negative $I^\mathrm{mod}_\mathrm{dc}$ (see supplementary material for analysis with varying $\mu_0H$).
\begin{figure}
	\includegraphics[]{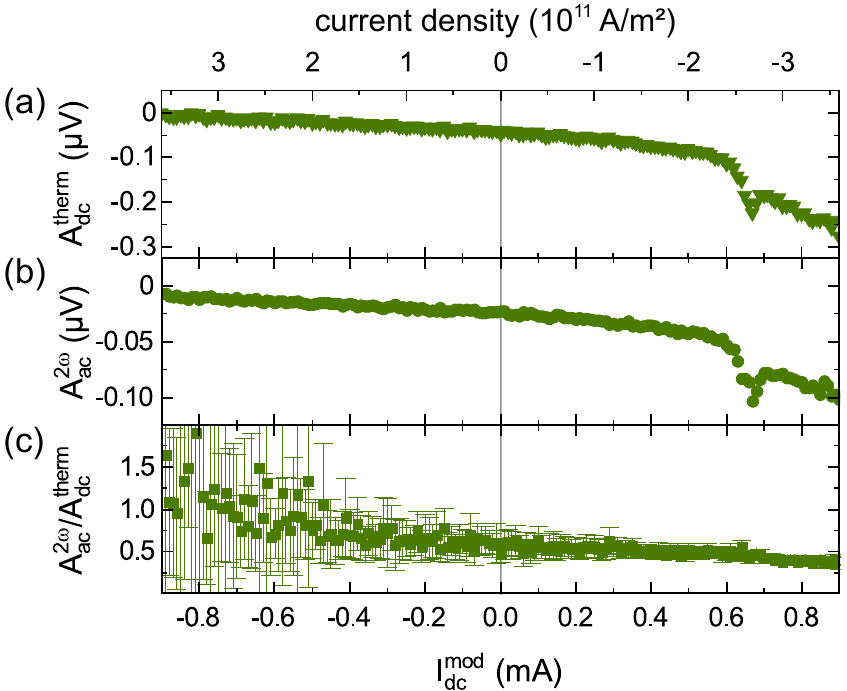}
	\caption[]{Extracted amplitudes (a) $A^\mathrm{therm}_\mathrm{dc}$ and (b) $A_\mathrm{ac}^\mathrm{2\omega}$ for $\mu_0H=\SI{60}{\milli\tesla}$  (as indicated in Fig.~\ref{fig:fig2admr}) of the thermally injected magnon transport signal for the dc and  the ac scheme versus the dc charge current $I^\mathrm{mod}_\mathrm{dc}$. (c) Ratio of the extracted amplitudes for the ac and dc configuration. Over the whole modulator current range the ratio shows a nearly constant behavior ($ A_\mathrm{ac}^\mathrm{2\omega}/A^\mathrm{therm}_\mathrm{dc}\simeq0.5$).}
	\label{fig:fig6currsweep200nm500nmtherm}
\end{figure}

In similar fashion, we extract the amplitudes $A^\mathrm{therm}_\mathrm{dc}(\pm\mu_0H)$ and $A_\mathrm{ac}^\mathrm{2\omega}(\pm\mu_0H)$ of the thermally injected magnons as a function of $I_\mathrm{dc}^\mathrm{mod}$ for the same magnetic field magnitude of $\mu_0H=\SI{60}{\milli\tesla}$. The results are shown in Fig.~\ref{fig:fig6currsweep200nm500nmtherm}(a) and (b) for the dc and ac configuration, respectively. The qualitative dependence on $I^\mathrm{mod}_\mathrm{dc}$ is identical for $A^\mathrm{therm}_\mathrm{dc}$ and $A^\mathrm{2\omega}_\mathrm{ac}$ for all current ranges. In agreement with previous reports~\cite{Wimmer2019spin}, we find a significant kink in  $A_\mathrm{dc}^\mathrm{therm}$ and $A_\mathrm{ac}^\mathrm{2\omega}$ above a certain critical current value. To account for the differences of the absolute amplitude values, we calculate the ratio  $A_\mathrm{ac}^\mathrm{2\omega}/A_\mathrm{dc}^\mathrm{therm}$, shown in Fig.~\ref{fig:fig6currsweep200nm500nmtherm}(c). The ratio is nearly constant over the whole modulator current range and has a value of 0.5 within the experimental error for all measured magnetic field magnitudes (see supplemental information for other $\mu_0H$).  
The small deviation, most notably in the negative bias regime, may be explained by the low thermal signal amplitude in our devices (yielding a worse signal-to-noise ratio) and differences in thermal landscape due to a difference in the average applied heating power for ac and dc measurements.
Nevertheless, the thermally generated signals nicely agree with our simple model of the detector voltage signal. However, the quantitative comparison of the thermal signal is not suitable to detect higher order contributions. 

In summary, we compared two measurement techniques, both allowing for an all-electrical generation and detection of pure spin currents in MOI/HM heterostructures.
On the one hand, we employ a dc-detection technique, where we utilized a modified current reversal method to take into account the modulator in a three-terminal nanostructure and to differentiate between SHE and thermally injected magnons 
arriving from the injector at the detector. 
On the other hand, we used an ac-readout technique, where lock-in detection of the first and second harmonic signal is utilized to distinguish between these two magnon contributions.  We demonstrate that the dc and ac  technique are both well suited to investigate incoherent magnon transport in these three-terminal structures. In addition, our results show that below a critical $I^\mathrm{mod}_\mathrm{dc}$ the detector voltage has contributions linear and quadratic in $I^\mathrm{inj}$. This especially manifests itself as a full quantitative agreement between $V^\mathrm{SHE}_\mathrm{dc}$ and $V^\mathrm{1\omega}_\mathrm{ac}$, which allows to compare results obtained with different techniques with higher confidence. For large modulator currents, deviations are observed, indicating a contribution of higher order in $I^\mathrm{inj}$ to the detector voltage. This sheds new light onto this nonlinear contributions appearing above a certain threshold value (corresponding to the damping compensation regime in our previous work).~\cite{Wimmer2019spin}

\section*{Supplementary Material}
See supplementary material for details on the fabrication process and the measurement techniques, separate measurements of an additional sample investigating the SHE injected magnons, angle-dependent measurements of the presented sample for negative filed polarity, a study of the field dependence of the extracted amplitudes of the electrically and thermally induced magnons for the dc- and the ac-detection technique and an investigation of the third harmonic voltage signal.

\begin{acknowledgments}
	
We gratefully acknowledge financial support from the Deutsche Forschungsgemeinschaft (DFG, German Research Foundation) under Germany's Excellence Strategy -- EXC-2111 -- 390814868 and project AL2110/2-1.
	
\end{acknowledgments}

\section*{Data Availability}
The data that support the findings of this study are available from the corresponding author upon reasonable request.



%

\end{document}